\documentclass[]{spie}  

\usepackage{amsmath,amsfonts,amssymb}
\usepackage{graphicx}
\usepackage[colorlinks=true, allcolors=blue]{hyperref}
\usepackage{caption}
\usepackage{subcaption}

\title{CCAT-prime: The 850~GHz camera for Prime-Cam on FYST}

\author[a,b,c,d]{Scott C. Chapman}
\author[d]{Anthony I. Huber}
\author[c]{Adrian K. Sinclair}
\author[e]{Jordan D. Wheeler}
\author[e]{Jason E. Austermann}
\author[e]{James  Beall}
\author[c]{James Burgoyne}
\author[f]{Steve K.\ Choi}
\author[f]{Abigail Crites}
\author[f]{Cody J.\ Duell}
\author[c]{Jesslyn Devina}
\author[e]{Jiansong Gao}
\author[i]{Mike Fich}
\author[b]{Doug Henke}
\author[g]{Terry Herter}
\author[b,d]{Doug Johnstone}
\author[b]{Lewis B.\ G.\ Knee}
\author[f,g]{Michael D.\ Niemack}
\author[h]{Kayla M. Rossi}
\author[g]{Gordon Stacey}
\author[c]{Joel Tsuchitori}
\author[e]{Joel Ullom}
\author[e]{Jeff Van Lanen}
\author[f]{Eve M. Vavagiakis}
\author[e]{Michael Vissers}
\author[ ]{the CCAT-prime collaboration}
\affil[a]{Dept. of Physics and Atmospheric Science, Dalhousie University, Halifax, Canada}
\affil[b]{NRC Herzberg Astronomy \& Astrophysics Research Centre, Victoria, Canada}
\affil[c]{Dept. of Physics and Astronomy, University of British Columbia, Vancouver, Canada}
\affil[d]{Dept. of Physics and Astronomy, University of Victoria, Victoria, Canada}
\affil[e]{National Institute of Standards and Technology, Boulder, Colorado, USA}
\affil[f]{Dept. of Physics, Cornell University, Ithaca, USA}
\affil[g]{Dept. of Astronomy, Cornell University, Ithaca, USA}
\affil[h]{Cornell Center for Astrophysics and Planetary Sciences, Cornell University, Ithaca, USA}
\affil[i]{Waterloo Centre for Astrophysics, University of Waterloo, Waterloo, ON, Canada}

\authorinfo{Further author information: (Send correspondence to SCC: E-mail: scott.chapman@dal.ca, Telephone: 1 778 678-7161)}

\begin{document} 
\maketitle

\begin{abstract}
The Fred Young Submillimeter Telescope (FYST) at the Cerro-Chajnantor Atacama Telescope prime (CCAT-prime) Facility will host Prime-Cam as a powerful, first generation camera 
with imaging polarimeters working at several wavelengths and spectroscopic instruments aimed at intensity mapping during the Epoch of Reionization. Here we introduce the 850 GHz (350 micron) instrument module. This will be the highest frequency module in Prime-Cam and the most novel for astronomical and cosmological surveys, taking full advantage of the atmospheric transparency at the high 5600 meter CCAT-prime siting on Cerro Chajnantor. The 850 GHz module will deploy $\sim$40,000 Kinetic Inductance Detectors (KIDs) with Silicon platelet feedhorn coupling (both fabricated at NIST), and will provide unprecedented broadband intensity and polarization measurement capabilities. The 850\,GHz module will be key to addressing pressing astrophysical questions regarding galaxy formation, Big Bang cosmology, and star formation within our own Galaxy. We present the motivation and overall design for the module, and initial laboratory characterization.


\end{abstract}

\keywords{CCAT-prime, CCAT, RFSoC, MKID, frequency multiplexed, kinetic inductance}

\section{INTRODUCTION}
\label{sec:intro}  

The CCAT-prime telescope is a 6-m aperture submillimeter (sub-mm) to millimeter (mm) wave telescope to be completed by 2024.  It is being built by an international consortium led by Cornell University, and including a large Canadian partnership led by five universities (Waterloo, Dalhousie, UBC, Toronto, Alberta). In particular, Canada is leading the development of the highest frequency, 850 GHz,  instrument for the telescope. CCAT-prime's extremely wide field of view (FoV) crossed-Dragone design will enable far faster mapping than existing facilities (Figure 1: [1]) 
The off-axis 6-m design achieves very low, $\sim1\%$, emissivity to take advantage of the Cerro Chajnantor site for sub-mm/mm wave astrophysics [2]. 
The high surface accuracy, half-wavefront error~$\leq$~10.7~$\mu$m, ensures excellent sub-mm sensitivity, crucial for observations at the highest frequencies. The mirror shapes are modified to reduce coma, providing an FoV of  2$^{\circ}$ in diameter at 850\,GHz (350$\mu$m) allowing for high Strehl optical designs for the module described in this paper. At lower frequencies, up to 7.8$^{\circ}$ FoV (at 100\,GHz) can be utilized. The large FoV enables the illumination of approximately ten times more detectors than the current generation of millimeter telescopes [3]. 
The CCAT-prime telescope is described in [4]. 
This telescope concept was adopted by the Simons Observatory (SO) in 2017, and development of the telescope for SO has proceeded in conjunction with CCAT-prime. Both telescopes are being built by Vertex Antennentechnik GmbH. {To ensure effective use of the large FoV, the telescope will house a seven module camera, Prime-Cam [5].} For Prime-Cam we have adopted cryogenic and optical designs from the Simons Observatory Large Aperture Telescope Receiver (LATR) (e.g.\ [6,7]).  

\begin{figure}[t!]
   \centering
   \includegraphics[width=1.0\textwidth]{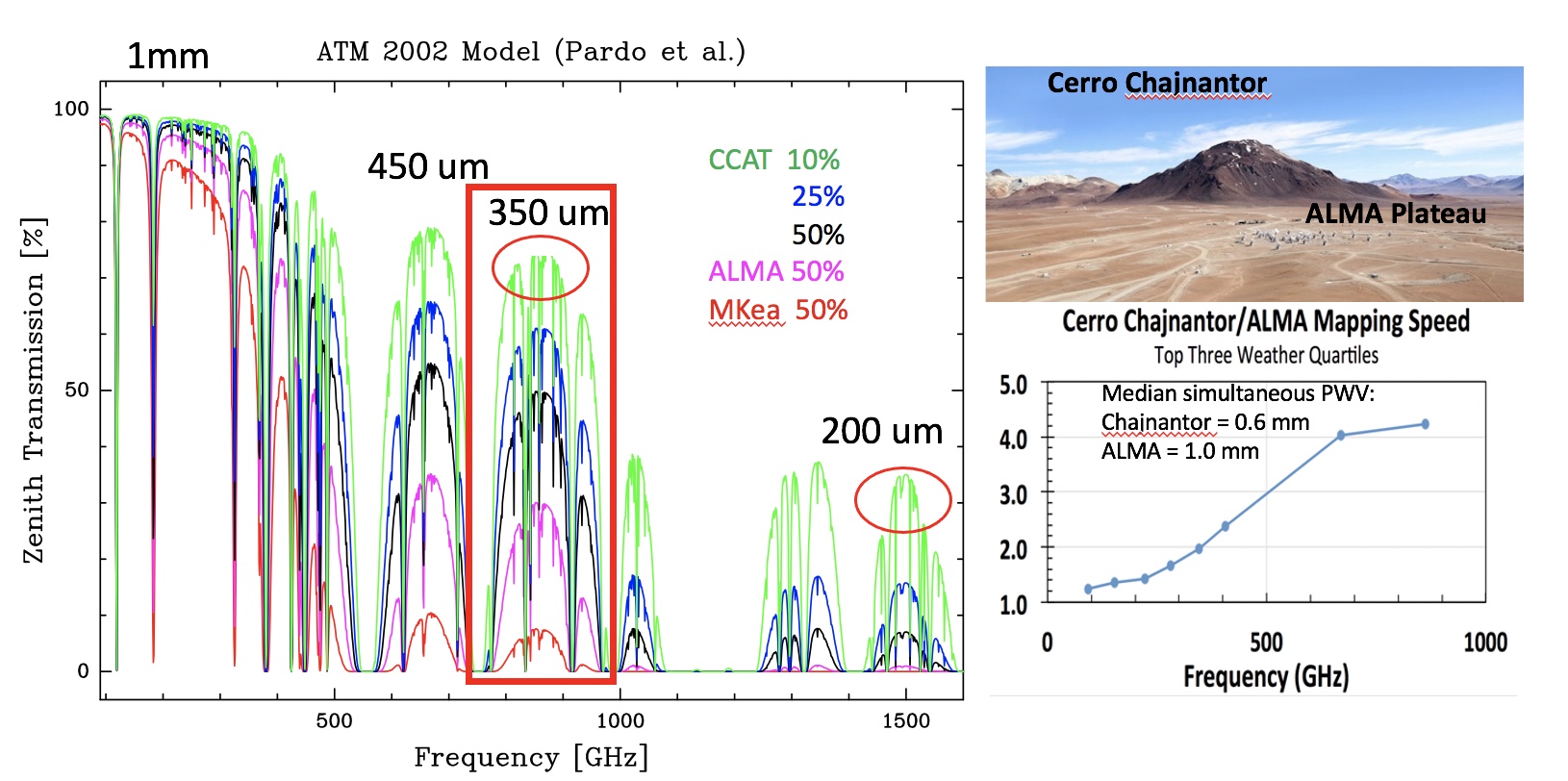} 
   \centering
   \caption{{\it Left}: Atmospheric transmission at the CCAT-prime site, Cerro Chajnantor, emphasizing the exquisite 850 GHz (350$\mu$m) transparency compared to the ALMA and Mauna Kea sites. Note that a future instrument will be able to take advantage of the 1500 GHz (200$\mu$m) window. {\it Right Top}: a photo of the site, and {\it Right Bottom}: the improvement in sub-mm mapping speed expected at 5600\,m relative to that of the ALMA Plateau [11, 10]. 
   }
  \label{figs2:CCATpcam}
\end{figure}

A single instrument module cryogenic receiver, Mod-Cam, is being built for first light observations with a 280 GHz instrument module on FYST [8], which will be followed by the larger seven module instrument Prime-Cam [5] which will enable unique observations that address astrophysical questions ranging from the physics of star formation to Big Bang cosmology. Prime-Cam will simultaneously cover five bands in five separate camera modules, spanning 190 to 900\,GHz (1.6 to 0.33\,mm), along with two spectroscopic modules.  The wavelength coverage, sensitivity, spatial resolution, and large field of view (FoV) of Prime-Cam on CCAT-prime allow for a set of wide-area surveys, between 5 and 15,000\,deg$^2$, to be conducted in order to address many scientific goals, described in [9]:

\begin{enumerate}
\itemsep 0em
 
 \item Directly trace the evolution of dusty-obscured star formation in galaxies since the epoch of galaxy assembly, starting $>$ 10\,billion years ago;  
 
  \item Trace the formation and large-scale three-dimensional clustering of the first star-forming galaxies during the Epoch of Reionization through wide-field, broadband spectroscopy; 
  
  \item Constrain dark energy and feedback mechanisms by measuring the physical properties and distribution of galaxy clusters via the Sunyaev-Zeldovich (SZ) effects on the CMB; 
  
  \item Enable more precise constraints on inflationary gravity waves and light relics by measuring polarized CMB foregrounds and Rayleigh scattering. 
  
  \item Monitor and search for time-dependent events: a wide variety of transient sources are becoming a large focus in (sub-)mm surveys;

\item  Determine the role of magnetic fields in star formation by measuring Galactic polarization.

\end{enumerate}

Enabled by the very low precipitable water vapor at the CCAT-prime site, the highest frequency camera module designed for the Prime-Cam instrument will perform 850\,GHz (350\,$\mu$m) continuum surveys utilizing kinetic inductor detectors (KIDs; [10,2]). 
All of the science goals above will leverage this novel wide field and polarization sensitive 850\,GHz module.

\subsection{Science Goals}

While the full CCAT-prime  capabilities are driven by a range of exciting science cases, four in particular are enabled by the high frequency 850 GHz (short wavelength 350$\mu$m) camera module described here. Further details can be found in [9].

\noindent{\bf How do galaxies form over Cosmic Time?}
Stars form from the collapse of molecular gas clouds. Much of this star formation is hidden from our view due to dust within the clouds that absorbs starlight, heats up, and reradiates the power at far-infrared wavelengths. Half of the starlight emitted through cosmic time is obscured by dust, which we measure as the cosmic infrared background (CIB). Many of the most interesting and pivotal objects for understanding galaxy formation in the early Universe are completely obscured by this dust, and only observable in the sub-mm bands [12]. Therefore, to understand the star formation history of the Universe, one must measure both 
redshifted optical emission from stars, 
and 
infrared emission from the dust which is redshifted into the sub-mm bands.  About 80\% of the CIB 
at redshifts $\sim1-3$ has been resolved into individual galaxies by the {\it Spitzer} and {\it Herschel} space telescopes. Yet at earlier times, crucial to the formation epoch of galaxies, the fraction resolved drops to just 10\%.

CCAT-prime will effectively reach up to three times deeper than the {\it Herschel} surveys, which are limited by source confusion, securely detecting hundreds of thousands of galaxies at redshifts as high as 7, less than 1\,Gyr after the Big Bang [9]. The 850 GHz band is crucial to pre-select the highest redshift candidates as very red or even drop outs. CCAT-prime will unveil about 40\% of the star formation at these epochs [13]. Its wide-field, $>$200 deg$^2$, surveys sample environments on large scales that cannot be mapped with interferometers like ALMA, and reach down to luminosities that remained inaccessible to {\it Herschel}. The 850 GHz band is especially novel and critical to extract physical properties, in particular infrared luminosities and dust-obscured star formation rates, to far greater precision than possible with lower frequency surveys  (e.g., the Large Millimetre Telescope in Mexico). 

Combining CCAT-prime surveys with synergistic work in the optical and near-infrared (e.g., DES, and the upcoming LSST, Euclid and WFIRST) will identify the key parameters that regulate star formation, such as environment and matter content, over cosmic time. 
The gas rich galaxies experiencing violent and short-lived starbursts in the early universe that created the local giant elliptical galaxies will be identified in statistical samples to reveal the star formation processes leading to ``normal'' galaxies like the Milky Way during the epoch of their assembly ($z\sim5$). The CCAT-prime wide-field survey promises to uncover the most intense starbursts in the universe, including those entirely missed by even the deepest optical and near-infrared surveys  (e.g., [12]). 
In total, $>$600,000 individual galaxies above a signal-to-noise ratio of 5 will be detected in the full survey, $>$1000 of which are expected to be at redshifts 5--8, in this largely unexplored period $\sim1$ billion years after the Big Bang.

\noindent{\bf What is the fundamental physics determining the early universe?} 
In recent years enormous progress has been made using Cosmic Microwave Background (CMB) temperature and polarization measurements to constrain cosmological parameters and characterize large scale structure. The CMB research community is developing plans for a next generation ``Stage IV'' CMB survey (CMB-S4) to achieve dramatic improvements in constraints on inflationary gravitational waves and light relic species (e.g., neutrinos). CCAT-prime is a potential telescope platform for CMB-S4, and it also offers unique capabilities for important advances in high-frequency polarization science before CMB-S4. In particular, the excellent high-elevation site and telescope capabilities will enable sensitive measurements in the telluric windows between 0.1 and 1 THz. No other site or telescope on Earth offers comparable sensitivity to CMB polarization and galaxy clusters spanning this entire frequency range, and no space or balloon telescopes offer comparable resolution. The proposed polarization sensitive 850 GHz camera, specifically, can play a unique and key role in this science [9]. These CMB cases require precise modelling of polarized dust properties in our Galaxy, the foreground to these cosmology experiments, for which the 850\,GHz module will provide a unique capability. At the same time, these polarization maps will allow for unprecedented insights into magnetic turbulence within the star formation process.

\noindent{\bf Do magnetic fields regulate star formation?}
Over the past decades, the field of sub-mm polarimetry has matured [14], using polarized emission from dust grains aligned with the local magnetic field to map out plane of sky magnetic fields in both individual molecular clouds 
and dense star-forming clumps. 
The Prime-Cam instrument on CCAT-prime will have $>$20 times better resolution than the Planck all-sky polarization maps. 
Prime-Cam has better predicted sensitivity than any other sub-mm polarimeter, including the balloon-borne polarimeter BLAST-TNG [15], and the 850\,GHz camera module offers unique constraints and the highest CCAT-prime angular resolution. 

CCAT-prime will survey dust polarisation throughout extremely young, low column density molecular clouds, directly observing the role magnetic fields play in regulating star formation. By statistically comparing polarization maps with synthetic polarization maps made from molecular cloud simulations, these observations will be used determine the relative importance of magnetic fields, gravity, and turbulence, for clouds of different masses and evolutionary states. 
Together with the lower frequency modules, the 850 GHz camera module will connect the cloud scale magnetic field structure to the small scales which ALMA will measure for protostellar envelopes and protoplanetary disks. Within the moecluar cloud, Prime-Cam will also disentangle the magnetic field topology as it relates to filaments and their star-forming cores.

\noindent{\bf How are protostars assembled?}
The first step in building a star within a molecular cloud, is the formation of localized, gravitationally bound prestellar core. Once the core becomes unstable to collapse, material falls toward the centre at free-fall velocities. Angular moment due to rotation of the core, however, provides a centrifugal barrier and thus the majority of the material that will eventually build the star is deflected into a circumstellar disk. Assembly of the star therefore requires physical dissipation and angular momentum transfer mechanisms within the disk, allowing for inward diffusion of mass and eventual accretion onto the central (proto)star. Theoretical models and observations of star-forming regions both suggest that the accretion process within the disk is episodic, driven by non-linear instabilities rather than steady flow [16]. 
Constraining the physical parameters associated with these accretion events is challenging as the earliest phases of mass assembly are enshrouded by the prenatal envelope and therefore hidden at optical and infrared wavelengths [17]. 
Only recently, with the advent of large area, sensitive, submillimetre cameras have monitoring campaigns become possible [18,19]. 

With the 850\,GHz module CCAT-prime will efficiently observe several nearby star-forming regions containing hundreds of protostars with a few week cadence and over many years, essential monitoring to catch the temporal structure of the episodic accretion [9]. Each epoch measures the change in brightness of the protostellar envelope, which is directly related to the time-variable release of gravitational energy as mass falls onto the protostar. The 850\,GHz module, observing close to the peak of the envelope spectral energy distribution, most accurately reflects this asccretion luminosity change, whereas Prime-Cam's lower frequency observations yield the weighted change in dust temperature within the envelope, a much weaker signal. As a survey instrument with a large Fov, Prime-Cam will provide an extremely efficient platform for time-domain sub-mm observations.

\begin{figure}
    \centering
    \includegraphics[width = 0.7\textwidth]{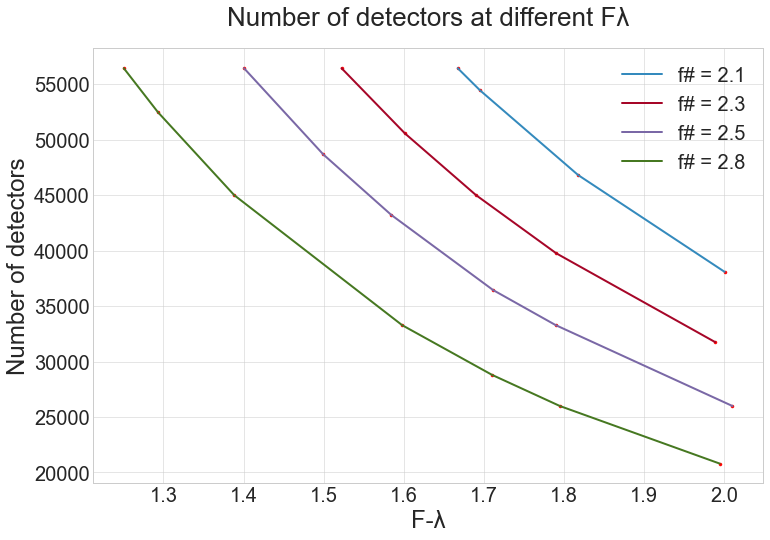}
    \caption{{\small Number of detectors required to fill candidate  camera designs with four different fields-of-view (f\#2.8 to f\#2.1, corresponding to 1.0$^{\circ}$ to 1.3$^{\circ}$ FoVs) as a function of F-$\lambda$ pixel spacing.
    The range of required detectors spans that which is feasible using single layer lithographic fabrication of the capacitors, inductors and feed lines (see section~3).}}
    \label{fig:detsvsfl}
\end{figure}

\begin{figure}
    \centering
    \vskip\baselineskip
    \begin{subfigure}{0.45\textwidth}
        \centering
        \includegraphics[width =\textwidth]{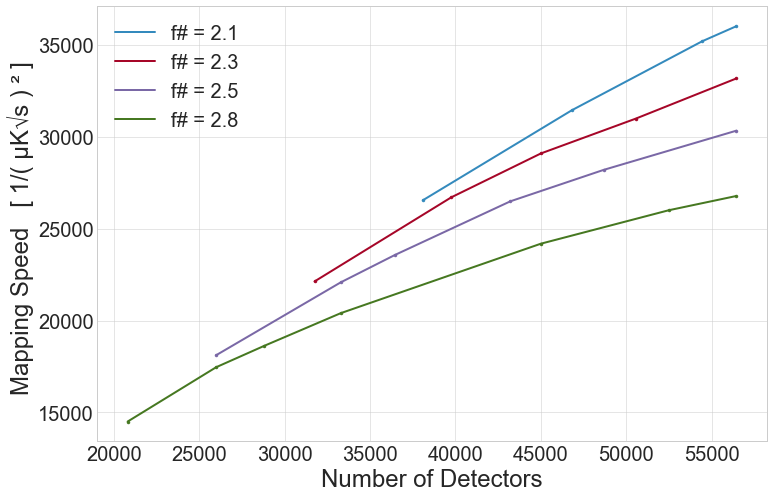}
        \caption[]
        {{\small Extended Sources}}
        \label{fig:MS Dets Extended}
    \end{subfigure}
    \hfill
    \begin{subfigure}{0.45\textwidth}
        \includegraphics[width=\textwidth]{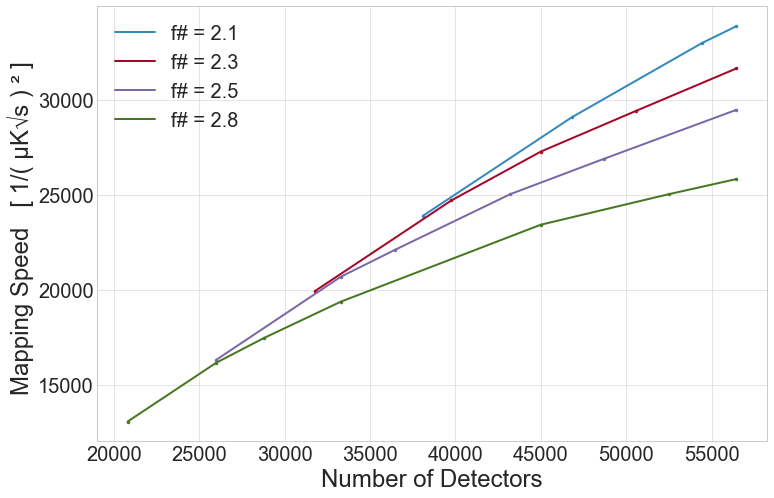}
        \caption[]
        {{\small Point Sources}}
        \label{fig:MS Dets Point}
    \end{subfigure}
 \vskip0.2cm
 \caption[]{\small Mapping speed as function of number of detectors for the same four FoV optical designs shown in Fig.~2. In \ref{fig:MS Dets Extended}, the mapping speed is calculated for extended sources while in  \ref{fig:MS Dets Point}, it is calculated for point sources. 
 For a fixed detector number $>$40,000, there is a marked $\sim$10\% gain in mapping speed for each 0.1$^{\circ}$ gain in FoV diameter, due to the  advantageous gain on the spillover/aperture efficiency curves for horn-coupled detectors.
 As the number of detectors is practically limited by the readout to $\sim$40,000 (Sinclair et al.\ 2022), a gain in field size provides a simple solution to mapping speed improvement. However, once anti-reflection losses from the addition of a 5th lens are factored in, the gain from increasing field size is largely removed.}
    \label{fig:MappingSpeedGraphs}
\end{figure}

\section{Instrument design and characterization}

\subsection{Mapping speed: building on submm mapping legacies} %

The CCAT-p 850\,GHz camera module builds on the legacy of the JCMT SCUBA and SCUBA-2 instruments for deep mapping of high Galactic latitude fields. While SCUBA-2 was able to survey $\sim$10 deg$^2$ at 350 GHz (850$\mu$m) [20] down to near the confusion limit ($\sim1\,$mJy for the 15$''$ beam), at  650 GHz (450$\mu$m), deep surveys of only 0.1~deg$^2$  were achieved, primarily from the Large Program ‘STUDIES’ [21] and Cosmology Legacy Survey  [22]. Because of the poor aperture efficiency of the JCMT dish at 450$\mu$m, the effective beam has a FWHM of 
$\approx10''$ [22], and survey confusion limit also approached 1\,mJy RMS [21].

At 850 GHz, the 1.3$^{\circ}$ FoV of the individual CCAT-p camera modules coupled with the atmospheric transmission advantage at 5600m (Fig.~1), implies more than a factor 1000$\times$ improvement in mapping speed achievable compared to a similar instrument on the JCMT, given the lower and wetter site on Mauna Kea and the smaller 8$'$ field of view. CCAT-p Prime-Cam 850 GHz (350$\mu$m) surveys will cover $>100$ deg$^2$ with a 15$''$ beam to comparable depths achieved by deep legacy surveys with SCUBA-2 at 450$\mu$m.
Furthermore, the 10$''$ JCMT beam at 450$\mu$m is not appreciably different than the CCAT-p beam  at 350$\mu$m (850 GHz), thus the confusion limits will be similar [9].
CCAT-Prime will thus be the preeminent sub-mm survey facility over the next decade.

\subsection{850 GHz module design considerations}

The baseline design for the 850\,Ghz module is to target illuminating roughly 5.5-m of the 6.0-m aperture Fred Young Submm Telescope, which provides f/2.6 at the telescope focus [23]. 
The approximately 0.4-m diameter optics tube illuminates three 150-mm detector wafers. The optics tube size provides an unobstructed 1.3$^{\circ}$ diameter FoV and keeps the size of the entrance window manageable. 

Huber et al.\ [24] present optical design considerations for the 850\,GHz camera module. 
They consider the trade-off space between 4 and 5 lenses and additional length of the optical design, which adds complexity due to the somewhat reduced space for the cryogenic wiring from the detector stage. The instrument module optics design is optimized for a telecentric, or close to telecentric, diffraction-limited image with minimal ellipticity, enabling polarization measurements across as much of the 1.3$^{\circ}$ diameter FoV as possible. Designs for 1.0$^{\circ}$ through 1.3$^{\circ}$ FoV can be delivered with good Strehl ratios ($>0.8$), but with increasing trade-off costs with larger FoV. A break point exists for the 4-lens design at $\sim1.1^{\circ}$, where for any larger a FoV, a more complex design must be considered and/or a 5th lens must be added.

We thus consider here FoVs of 1$^{\circ}$, 1.1$^{\circ}$, 1.2$^{\circ}$, and 1.3$^{\circ}$ (equivalently f\# 2.8, 2.5, 2.3, 2.1).
Fig.~2 shows the number of detectors required to fill the four candidate FoVs as a function of the F-$\lambda$ pixel spacing, with each pixel requiring two detectors in order to sample both polarizations. The key trade-offs are shown in Fig.~3 where the mapping speeds are assessed for these four field sizes as a function of the number of detectors filling the field.

For horn-coupled detectors, the formula below shows how the normalized sensitivity is affected predominantly by the spillover efficiency (for extended sources) or aperture efficiency (for point sources). The effects on the mapping speed are similar from both the spillover and aperture efficiencies (see Fig.~3).

$${\rm Sensitivity} \propto \frac{1}{\rm \sqrt{{ throughput \times \# detectors \times efficiency(spillover\ or\ aperture) }}}\ \ ,$$

$${\rm Mapping\ Speed }\propto \frac{1}{\rm ( Sensitivity)^2}\ \ .$$

\textbf{Measurement of sky intensity distribution (extended source):}
The signal power absorbed by a detector is produced by the astronomical sky brightness when observing extended sources.
The spillover efficiency of the horns is the dominant factor affecting mapping speed for different F-$\lambda$ spacings, and is dependent on the horn edge taper. 
The edge taper values for F-$\lambda$ between 1 and 2 are taken from [25].

\textbf{Observation of point sources in the field:}
The detector coupling efficiency to the point spread function is important when measuring the flux density of point sources. 
The realistic maximum feedhorn aperture efficiency is 0.7 for an aperture of 2\,F-$\lambda$. However increasing the density of horns (pixels) to $<$2\,F-$\lambda$ increases the overall mapping speed for a fixed FoV.
The aperture efficiency at each F-$\lambda$ is taken from [25].

For a fixed detector number of at least 40,000 KIDs, there is a marked $\sim$10\% gain in mapping speed for each 0.1$^{\circ}$ increase in FoV diameter due to the  advantageous gain on the spillover/aperture efficiency curves for horn-coupled detectors. Less than this number and the gain in going to fields larger than 1 degree is negligible (the detectors are just spread out at larger than 2 F-$\lambda$ spacing).
As the number of detectors is practically limited by the readout to $\sim$40,000 [26], a gain in field size provides a viable direction for mapping speed improvement.
 However, once anti-reflection losses from the addition of a 5th lens are factored in [24], the gain from increasing field size is substantially eliminated.
 Thus, an optimal design currently being considered is for a 4-lens, 1.1$^{\circ}$ FoV, with $\sim$40,000 KIDs and F-$\lambda \sim$1.65, divided between  3 hexagonal arrays.
Compared to the baseline of 20,000 KIDs and a 1$^{\circ}$ FoV, Fig.~3 shows a two times gain in mapping speed for this design.



\subsection{Kinetic Inductance Detectors (KIDs)} %

The 850\,GHz array will use  kinetic inductance detectors (KIDs), as with the other Prime-Cam modules operating at lower frequencies [e.g., 8,30]. As described in this work and references therein, the photon sensitivity of a KID results from a superconducting resonator that derives a significant fraction of its total inductance from the kinetic inductance of an absorbing strip (inductive in this case). 
Photons which interact with the absorbing element of the detector will break Cooper pairs creating quasi particles. This causes a measureable change in the inductance through the resulting shift in the resonant frequency and quality factor (width of the resonance peak).

Fabrication of the first 850\,GHz KID test device was recently completed by the Quantum Sensors Group at the National Institute of Standards and Technology (NIST) in Boulder, CO. The 850\,GHz KIDs have benefited from the experience gained through  detector developments for the BLAST-TNG [15] and TolTEC [27] receivers.
For these broadband 850 GHz detectors, the change in optical power for each polarization is measured separately for each pixel.
The KIDs are lithographically defined superconducting microwave components separated in a single pixel into the two polarizations via different microstrip lines, meander inductors and inter-digitated capacitors. 

\begin{figure}
    \centering
    \includegraphics[width = 0.9\textwidth]{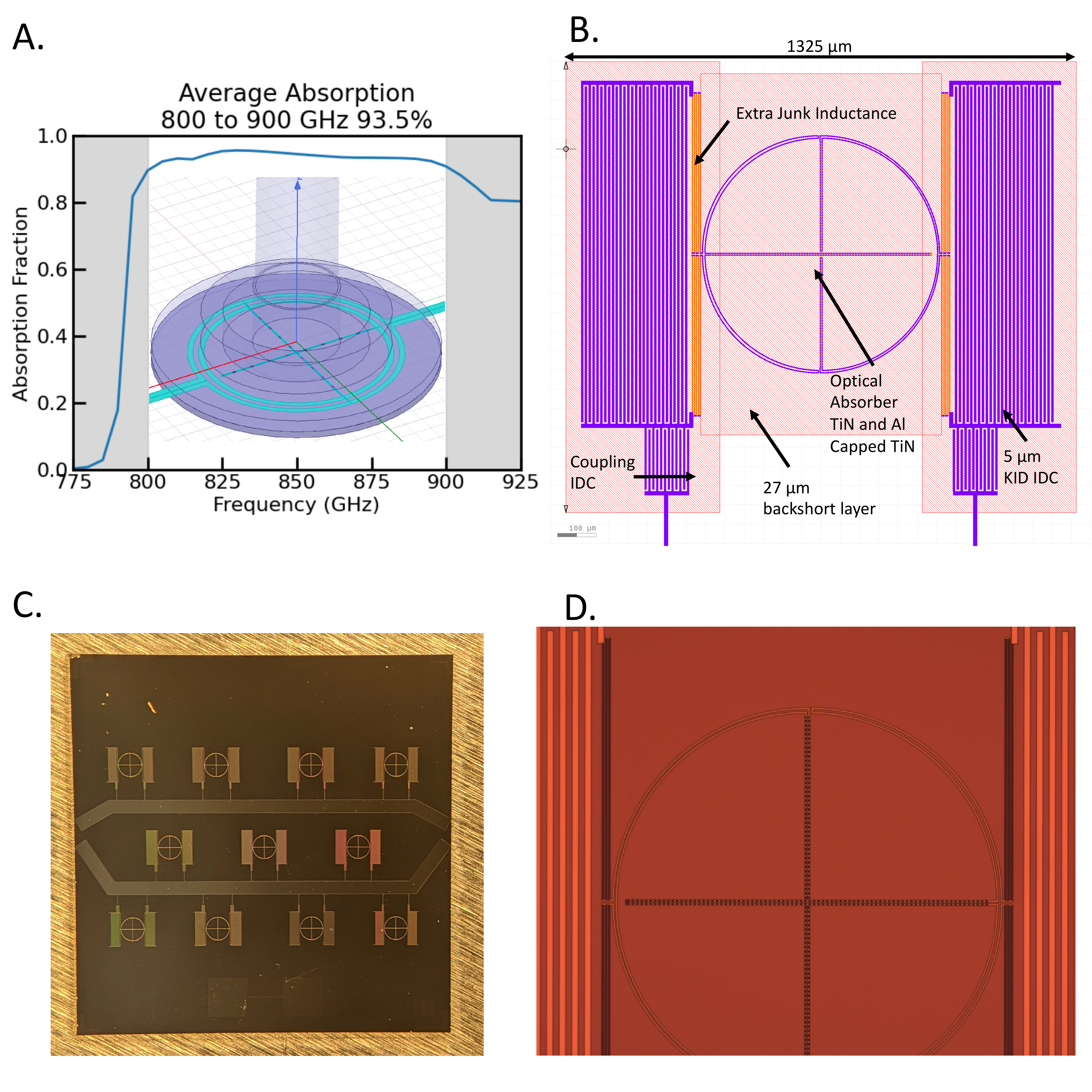}
    \caption{{\small A. preliminary optical coupling design and performance for KIDs. B. A test KID device designed to test adding extra inductance to relax demands on capacitor finger widths. C. The full fabricated test device with 11 pixels each with different amounts of extra "Junk" inductance or different capacitor finger widths. D. A microscope image of the pixel shown in B. }}
    \label{fig:KID}
\end{figure}

For KIDs in the 850 GHz camera, TiN with low T$_c$,
800 to 1000 mK, is targeted. A decrease in T$_c$ results in improved 1/f inductor noise for TiN KIDs pushing the 1/f knee to lower frequencies in an effort to make detector 1/f noise sub-dominant to the atmospheric 1/f noise. Preliminary absorber and waveguide interface optimization shows that 93\% detector efficiency can be obtained between 800 and 900 GHz (Figure \ref{fig:KID} A).

\begin{figure}
    \centering
    \includegraphics[width = 0.6\textwidth]{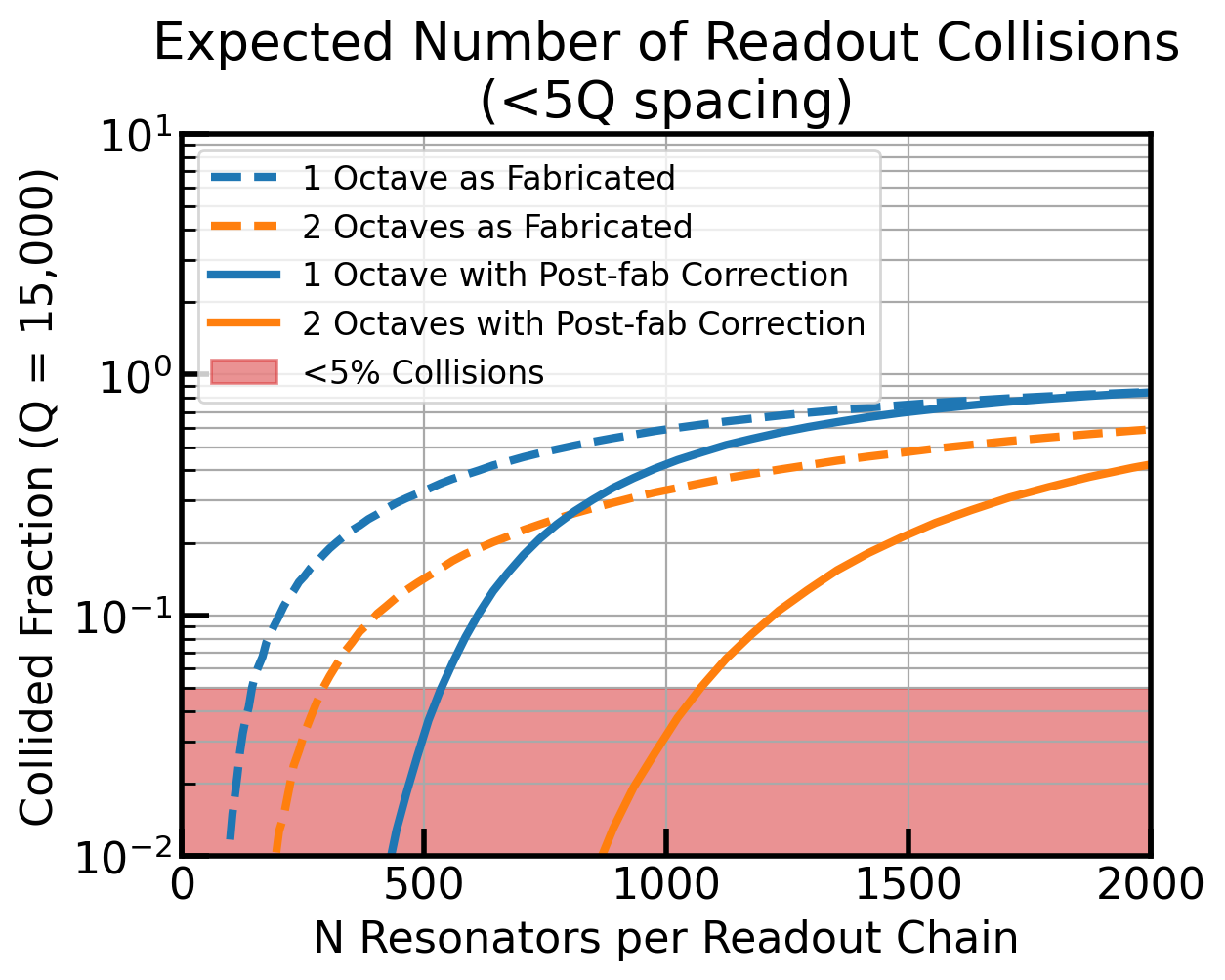}
    \caption{{\small Simulated number of readout collisions with and without post-fabrication resonant frequency correction for one and two octaves of readout bandwidth.}}
    \label{fig:collisions}
\end{figure}

Increased bandwidth microwave networks with larger network multiplexing factors are being pursued to enable greater detector counts per wafer while keeping readout infrastructure reasonable. This is being accomplished in two ways: First, by increasing the digital readout electronics bandwidth from 500 MHz to 1 GHz. Second, by changing the frequency scheduling of the detectors such that 1000 KIDs are placed within two octaves of readout bandwidth from 325 MHz to 1325 MHz. By increasing the multiplexing per microwave network by a factor of ~2 while also increasing the octaves of microwave network bandwidth from 1 to 2, the same dimensionless spacing between resonators in readout space as past arrays [28] can be maintained. This allows for maintaining a low number of collisions ($\sim$5\% after post-fab capacitor adjustment [29]) in readout space while achieving the required higher multiplexing factors (Figure \ref{fig:collisions}). 

An increased detector count per wafer presents design obstacles that need to be overcome,  primarily how to handle the decreased pixel size. The CCAT-p 280 GHz TiN array [30] has a pixel spacing of 2.75 mm, whereas the 850 GHz arrays desire a pixel pitch closer to 1.4 mm, almost a factor of 4 less in area. Scaling the capacitor by a factor of 4 in area would require capacitor finger widths a factor of 4 smaller as well. Smaller capacitor finger widths will lead to an increase in TLS Noise [15]. To keep the capacitor finger width similar to past arrays that show good TLS noise performance, extra inductance (junk inductance) can be added in by incorporating more squares of inductor into the KID design away from the optical absorbing part of the inductor. Increasing the inductance also enables pushing the lowest frequency resonators down from 500 MHz to 325 MHz. More inductance results in increased volume which will result in decreased sensitivity. For the 850 GHz camera module, however, this is actually advantageous because the loading level is higher than in past similar arrays [15]. The lower T$_c$ also has an effect of increasing detector sensitivity and thus requires additional volume to dilute the response such that the internal quality factors (Q$_i$ -- essentially the width of the resonances) are high enough under the appropriate loading levels. As junk inductance is increased, care must be taken that parasitic resonances are not driven down in frequency such that they are within the 325 to 1325 MHz readout network.
 
\begin{figure}
    \centering
    \includegraphics[width = 1.0\textwidth]{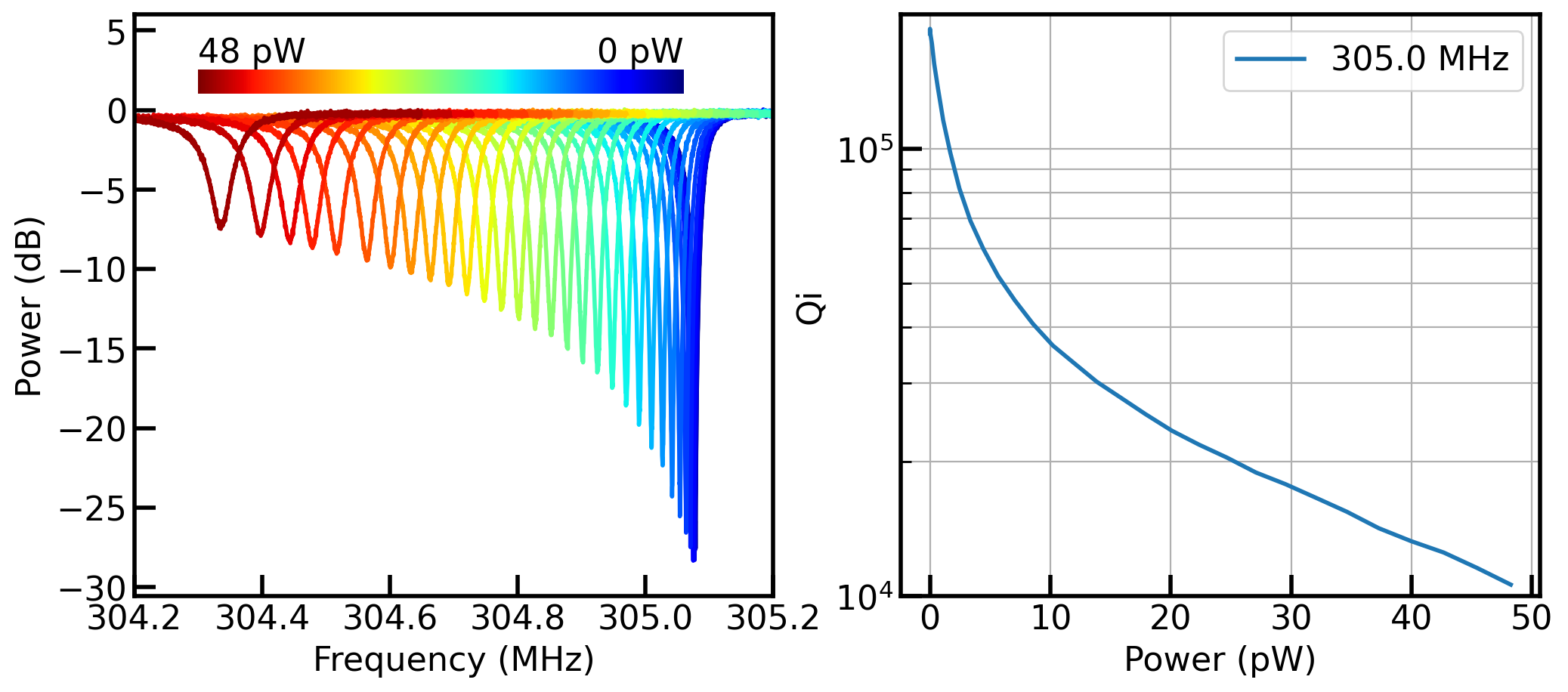}
    \caption{{\small Results of a cryogenic blackbody power sweep on the test resonator shown in Figure \ref{fig:KID} B.
    The blackbody temperature spans a range of 3\,K to 35\,K corresponding to optical powers from 0 to 48 pW given the optical filtering. An S$_{21}$ sweep of the resonator taken at varying optical load with a readout power of approximately -115 dBm is shown on the left. As optical power increases the frequency of the resonator decreases as does the internal quality factor (Q$_i$). The measured internal quality factor versus optical power is then shown on the right. 
    }}
    \label{fig:BB_sweep}
\end{figure}

To check that a 1.4 mm pixel pitch device can be made with 5$\mu$m capacitor finger widths with acceptable sensitivity and parasitic resonances, a preliminary test device was fabricated and tested (Figure \ref{fig:KID} B). These test devices showed that indeed a 325 MHz detector can be made with 5$\mu$m wide capacitor finger widths and with the first parasitic resonance occurring around 1600 MHz. Presumably, this is some sort of microstrip mode and thus puts a limit on how long the inductor can be. Therefore, to further reduce responsivity, increased linewidth will be required to increase the volume rather than increasing inductor length. For the preliminary test devices, an internal Q$_i$ of 10,000 was found at a loading level 50 pW (Figure \ref{fig:BB_sweep}) and thus further reduction in sensitivity is required to increase Q$_i$ to the desired 20,000 at the appropriate loading levels of 70 to 110 pW. This will be accomplished in future test devices by increasing the line width of the absorber. 

\subsection{Feed horns and interfaces} %

For optical coupling to the KIDs, gold plated, silicon stack feed horns are being baselined for the 850\,GHz module [31]. The lithographically produced design offers ease of scaling to large horn numbers, and a uniformity of process that yields improved cross-polarization performance. Fig.~7 shows an exploded view of typical lower frequency (150 GHz), silicon platelet stacks  (pre-metalization) already produced for other instruments. 
Individual Si wafer platelets are shown in the schematic cross-section of each horn design in Fig.~8. DRIE through-etching is used to fabricate the array of holes in each layer, and the ensemble is then glued together using lithographically defined alignment pins, and then gold plated.

\begin{figure}
    \centering
    \includegraphics[width = 0.33\textwidth]{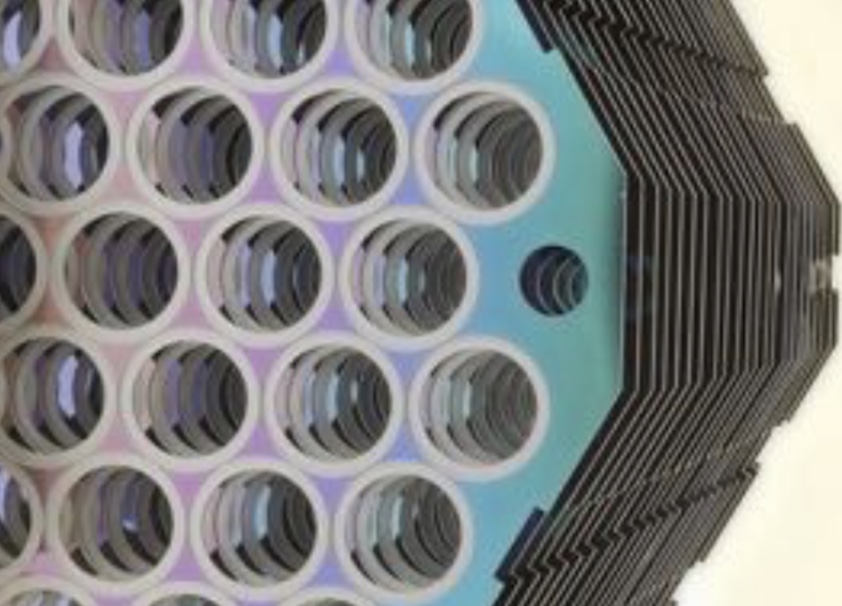}
        \includegraphics[width = 0.3\textwidth]{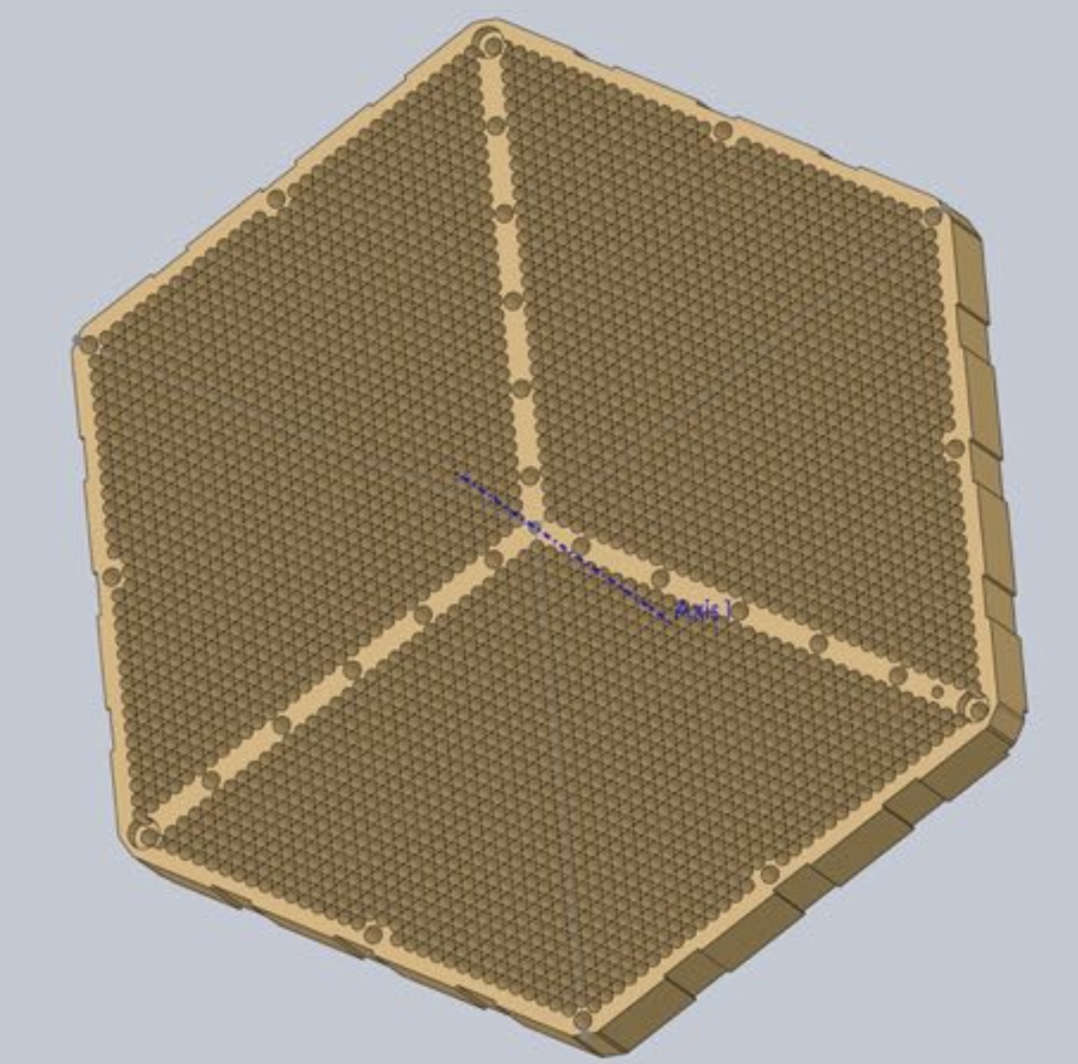}
    \includegraphics[width = 0.34\textwidth]{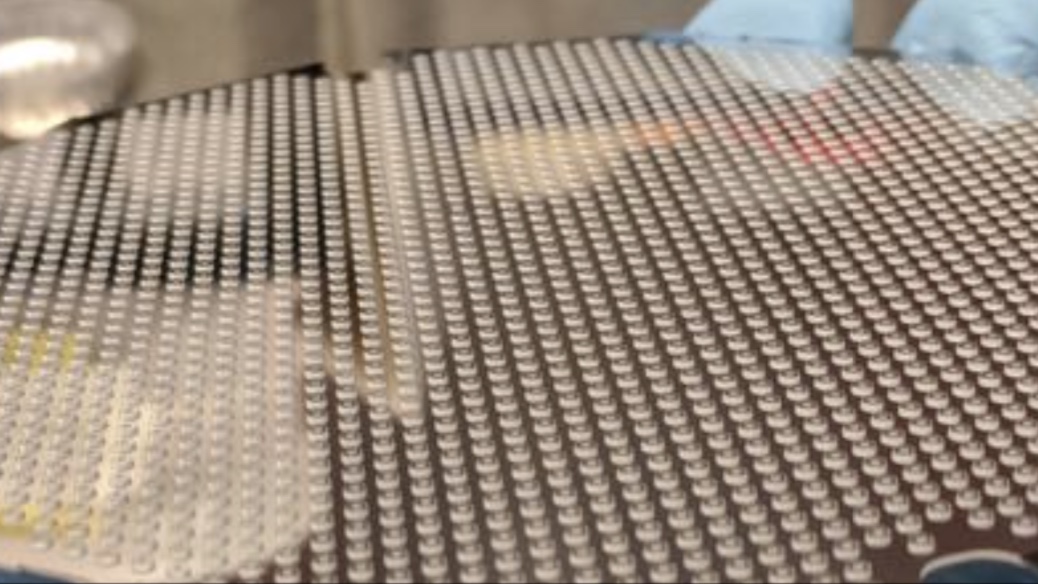}
        \includegraphics[width = 0.44\textwidth]{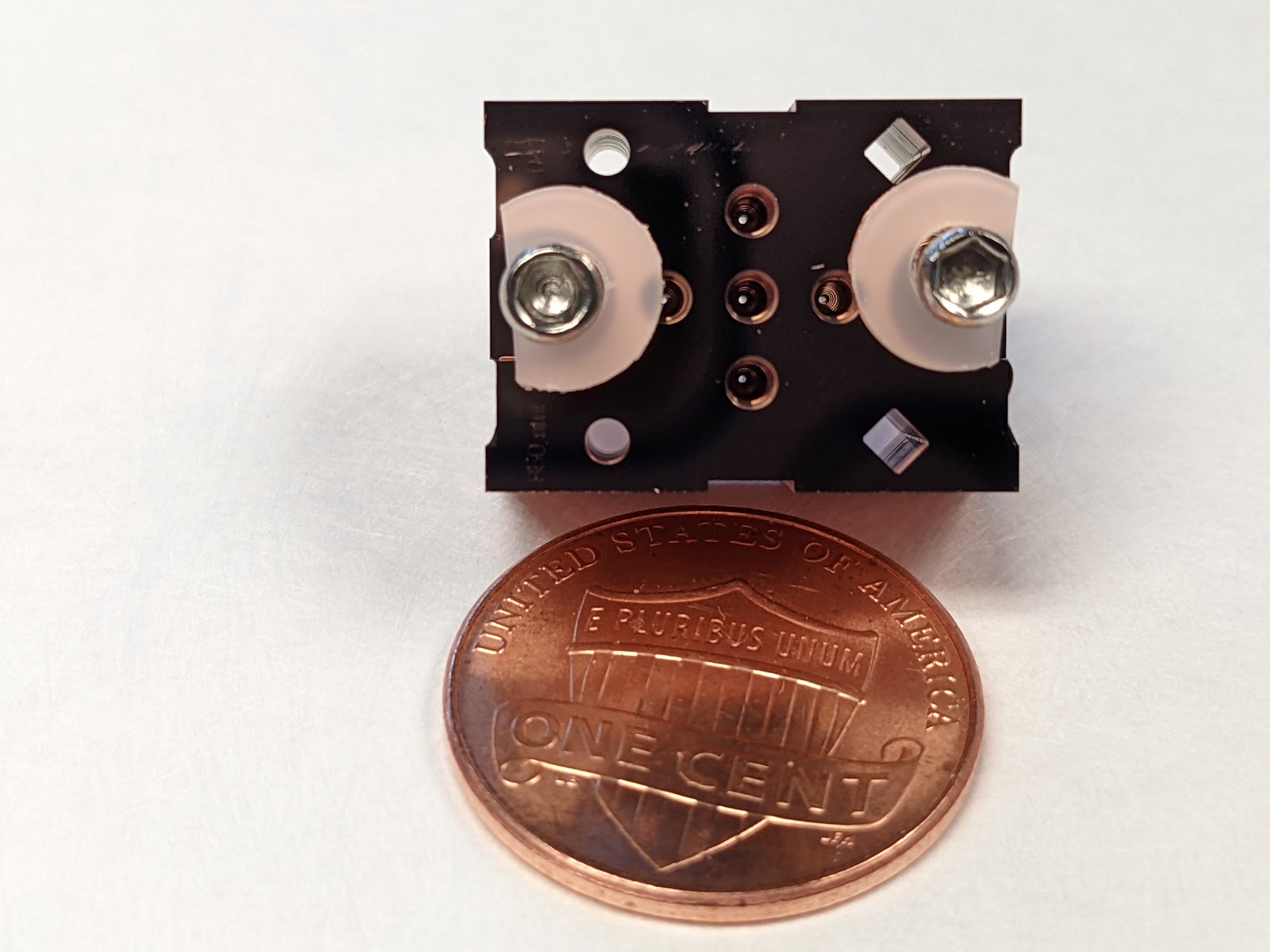}
            \includegraphics[width = 0.44\textwidth]{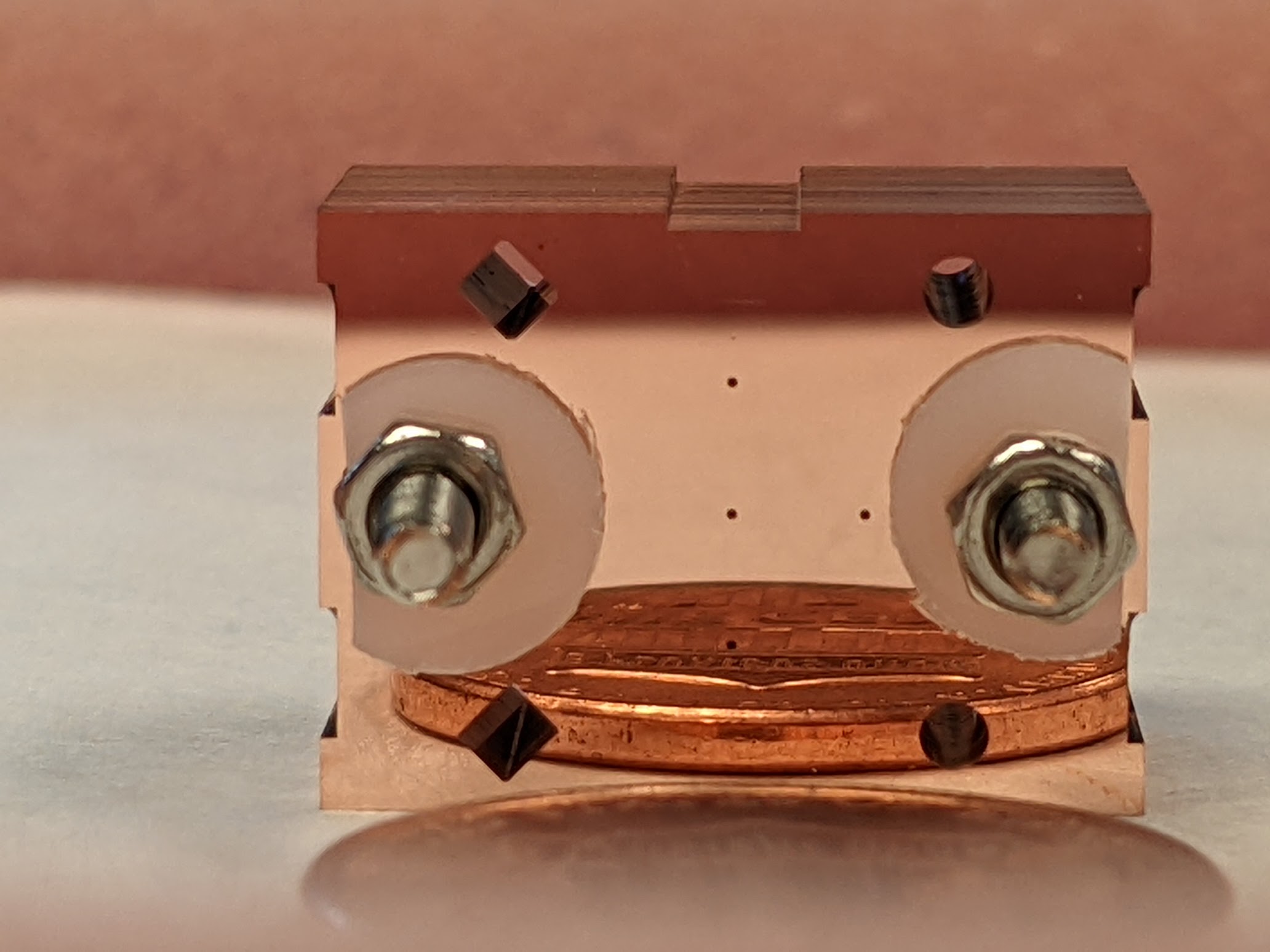}
 \vskip0.2cm
 \caption{{\small {\it Top Left:} Example Silicon platelet horn stacks  (pre-metalization) for a 150GHz array, produced using DRIE etch [31]. {\it Top Middle:} Example final horn array with a 2mm pixel pitch. The 850\,GHz baseline above will aim for a 1.4mm pitch. {\it Top Right:} Choke and Wave Guide structures: an SOI wafer sets the boss feature height and ring choke depth.
 {\it Bottom Row}
 Photos of the precision stacked prototype array of 5 feedhorns at 850\,GHz (entrance of horn on left, exit of horn on right).   
These are seed layered with copper (each platelet) and will be gold plated, which is the largest open question for typical processes translating to these higher frequencies (as the  waveguide is much smaller for gold to get down).  
    }}
    \label{fig:horn}
\end{figure}

Photos of the precision stacked prototype array of 5 feedhorns at 850\,GHz are also shown in Fig.~7.   
The prototypes are seed layered with copper (each platelet) and will be gold plated, which is the largest open question for typical processes translating to these higher frequencies (as the  waveguide is much smaller for gold to get down).  
Fig.~7 also illustrates the interface layers (between the horns and detectors) that will be fabricated for 850\,GHz, the choke and wave guide structures, where an SOI wafer sets the boss feature height and ring choke depths.

Fig.~8 shows designs and performance of 2.0mm and 1.4mm horns, spanning the range from our baseline to proposed instrument for 850\,GHz observations. For a 2.0mm pitch, Fig.~7 shows the Hex with a 33$\times$33 Rhombi pattern resulting in 3267 pixels (F-$\lambda \sim$2 for a 1$^{\circ}$ FoV), equivalent to 6534 detectors, or 19,602 detectors over the 3 Hex arrays. For a 1.4mm pitch, 39,204 detectors in 3 arrays provide F-$\lambda \sim$1.7 given a 1.1$^{\circ}$ FoV. 

The optimization of the 850\,GHz feedhorn profile is similar to that described in [15,27] for Al horns, and considers the beam coupling within 10.5$^{\circ}$, the cross-pol beam averaged by area within 10.5$^{\circ}$, and the beam asymmetry fraction within 10.5$^{\circ}$  (Fig.~8). This optmization leads to the final beam S11 throughput versus frequency for these designs.

\begin{figure}
\vskip0.5cm
    \centering
    \includegraphics[width = 1.0\textwidth]{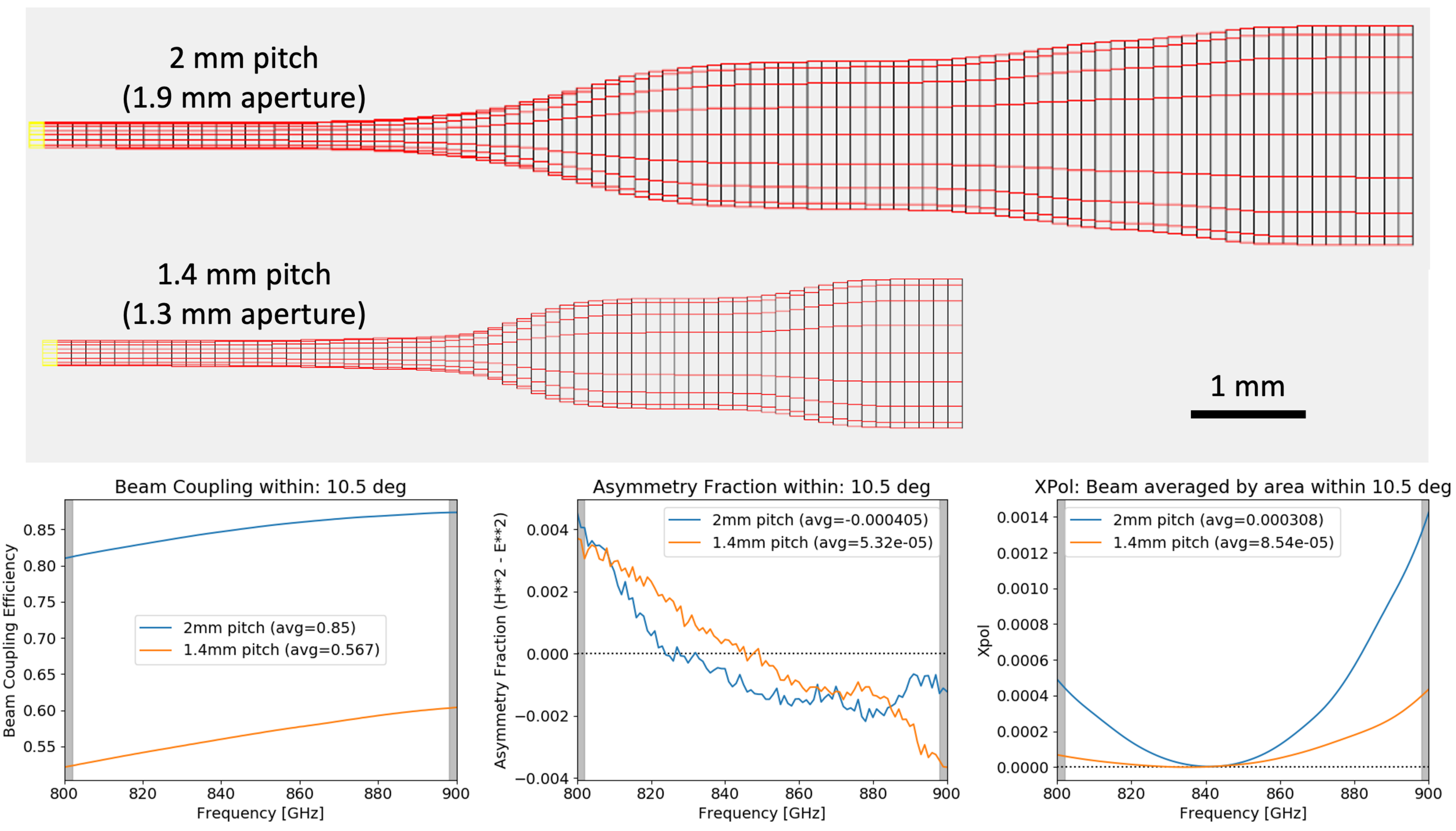}
 \vskip0.2cm
 \caption{{\small {\bf Top:} Designs and performance of 2.0~mm and 1.4~mm pitch platelet feedhorns, spanning the range from our baseline ($\sim$20,000 KIDs) to proposed ($\sim$40,000 KIDs) instrument for 850\,GHz observations. 
 Individual Si wafer platelets are shown in the schematic cross-section of each horn design. DRIE through-etching is used to fabricate the array of holes in each layer, and the ensemble is then glued together using lithographically defined alignment pins.
 {\bf Bottom row:}
 The optimizing of the 850\,GHz feedhorn profile considers the beam coupling, the beam asymmetry fraction, the induced cross-pol (Xpol), and reflections (S11) for the portion of the projected feedhorn beam profile that clears the cold stop aperture (approximately 10.5 degree half angle).  These metrics are shown in the bottom row of plots vs frequency within our band, except for reflection which is less than -25~dB at all points within the band. Gray shaded regions denote frequencies outside our nominal band.  These horn profiles were optimized for a platelet thickness of 250~$\mu$m thick wafers and 125~$\mu$m etch depths (i.e. two steps per wafer).  Further optimization and simplification of the horn profile will likely be pursued before full array production
    }}
    \label{fig:horndesign}
\end{figure}

\subsection{Readout} %
A 40,000 detector focal plane at sub-mm wavelengths is unprecedented, and ultimately is limited not by our ability to fabricate and cool these detectors, but by the need to read them out with a tractable system. 
Frequency domain readout has been used with success on TES detectors in many experiments, with multiplexing exceeding 60 times [32], whereby LC resonators are fabricated specifically to select specific TESs in the array for readout [33]. However the order of magnitude higher level of multiplexing required for large KID arrays ($>500\times$) is achievable entirely because the resonators become the detectors themselves in KID arrays (e.g, [27]).
As shown in Fig.~3, and discussed in Sec.~2.2, considering only the limitations of detector fabrication, trace width and effects on noise performance, coupled with the desire to maximize mapping speed, a 52,000 KID focal plane was initially considered. The practicalities of readout for this many detectors, running sufficient cryogenic RF lines for even 1000 KIDs per network, led us to settle on the 40,000 KID design (essentially resulting in a doubling of the cryogenic RF design being deployed in the ModCam receiver [30,8] and the other first light Prime-Cam modules [5].

To efficiently read out the $\sim$ 40,000 detectors, a microwave frequency multiplexed readout is being developed targeting the commercially available Xilinx Radio Frequency System on Chip (RFSoC) platform. The RFSoC integrates high speed digitizers, a re-programmable logic fabric, and two microprocessors onto a single chip, affording significant power savings typical of the far more expensive application specific integrated circuits (ASICs). Detector derived requirements and the overall design plan for RFSoC-based readout are described in [26]. A baseline design for Prime-Cam will be capable of reading out four independent RF networks each with $\sim$ 1000 detectors for $\sim$ 30W. The digital signal processing gateware draws its heritage directly from the BLAST-TNG design [34], 
which implements a frequency comb look-up table in external memory and utilizes two stages of channelization via a polyphase-filter bank and complex down-conversion with accumulation. Preliminary measurements of the  prototype system are also presented by [26] and show that the ZCU111 Gen 1 RFSoC development board will meet the expected bandwidth and noise requirements.



\subsection{Optics}

The modularity of Prime-Cam enables each optical path to be independently optimized for the science requirements of each instrument module. 
While the lower frequency modules adopt similar optical designs and layouts to the the Simons Observatory LATR [6,7], as described in [8], the 850\,GHz module represents a significant departure from these designs [24].
The 850\,GHz module contains a series of anti-reflection coated silicon optics, blocking filters, Lyot stop, and the detector focal plane. The module optics are designed to provide diffraction-limited (or near diffraction-limited) image quality across a wide FoV. 
An approximately 0.4-m diameter optics tube,  illuminates three 150-mm detector wafers. 
A ultra-high-molecular-weight polyethylene (UHMWPE) vacuum window and a series of approximately 6 cooled metal mesh infrared blocking and low-pass filters minimize emission and block undesired radiation in the instrument. A required alumina absorption filter represents particular challenges for the precision AR-coating features, and a laser ablation approach is being developed specifically for the 850\,GHz module. 
The telescope focus is transferred into the detector focal plane  by four refractive silicon lenses with metamaterial anti-reflection coatings. 
High resistivity silicon has extremely low loss (tan $\delta \sim$ few $10^{-5}$ at $T < 40$\,K, where $\delta$ is the loss angle), high thermal conductivity (ensuring lens temperature uniformity and limiting detector background loading), and a high index of refraction, $n \sim3.4$ [35]

\subsection{Cryogenics}

The 850\, GHz module will be hosted in the central position of the seven module Prime-Cam cryogenic receiver, where the Strehl ratio is high enough $>0.9$ to not further degrade the optical quality beyond that of the module reimaging system itself.
Within the vacuum shell of Prime-Cam, several cooling stages provide thermal isolation for the 850\,GHz optics tube and detector arrays. The temperature and stability requirements for the module  are derived from the requirements for the SO LATR [36,37]. 
80-K and 40-K temperature stages will hold optical filters. A short 80-K shield will be located at the front of the receiver. A 40-K shield will surround the interior of the receiver. A series of G10 tabs will support the assembly of thermal shields and provide thermal isolation as well as resilience against mechanical shocks and vibration. Inside the 40-K shield, the 850\,GHz module optics tube will be mounted on the 4-K temperature stage of the Prime-Cam cryostat. A 4-K shield will surround the back end of the optics tube. Cryomech PT-90 and PT-420 pulse tubes, each providing 90 W of cooling power at 80 K (PT-90), 55 W at 40 K (PT-420), and 2 W at 4 K (PT-420). A Bluefors LD-400 dilution refrigerator (DR) providing 400 $\mu$W of cooling power at 100 mK will be used to cool the 850\, GHz KID arrays to 100 mK and the final optical lenses to 1 K through thermal buses.
Around the exterior of the vacuum shell will lie supports for the room temperature  readout electronics cabinet hosting the nine ZCU111 RFSoC boards required for the 40,000 KIDs.

\section{Summary and Future Directions}

The Cerro­ Chajnantor Atacama Telescope prime (CCAT-p) Facility hosts the Fred Young Submillimeter Telescope (FYST), a wide-field, 6 meter aperture submillimeter telescope being built (completion in 2023) by an international consortium of universities led by Cornell University, including Canada as a major partner. Prime-Cam will be a powerful, first light camera for CCAT-p with imagers working at several wavelengths and spectroscopic instruments aimed at intensity mapping during the Epoch of Reionization. We presented the motivation and design of an instrument module in Prime-Cam, operating at 850\,GHz (350$\mu$m). This is the highest frequency on the instrument, and the most novel for astronomical surveys, taking full advantage of the atmospheric transparency at the high 5600 meter CCAT-p site on Cerro Chajnantor. This instrument will provide unprecedented broadband intensity and polarization measurement capabilities to address pressing astrophysical questions regarding galaxy formation, Big Bang cosmology, and star formation within our own Galaxy. 

The camera module focal plane is optimized for single frequency, polarization sensitive measurements at 850 GHz, achieving both a high mapping speed and point source sensitivity. The higher frequency, and resulting smaller pixel size than the other Prime-Cam modules, leads to a significantly different optimization from (for example) the 280\,GHz module. The availability of  densely populated arrays, combined with recent progress in demonstrating photon-noise-limited performance with KIDs coupled to silicon feedhorns, motivates the selection of KIDs for the 850-GHz focal plane.
In particular, around 13,000 KIDs with feedhorn coupling will be fabricated at NIST on a single 15-cm diameter wafer ($\approx$40,000 over 3 wafers in the focal plane), which is almost four times higher than the lower frequency arrays. This places large challenges on the frequency multiplexed readout system and cryogenic RF and feed throughs [24,26].  
We have presented the overall design for the camera module, detailing the KID detectors design and performance, the Silicon platelet feed horn design, and discussed initial laboratory characterization of the 850-GHz KID test devices. We referenced the optical and mechanical design of the module discussed in detail in [24], the required tolerances and performance metrics, the readout and  software directions detailed in [26]. 

The prototype KID arrays will be tested using a new Bluefors LD250 DR cryostat installed at the National Research Council, Herzberg Astronomy and Astrophysics center (NRC-HAA) in Victoria, BC. A cryogenic RF an custom LNA is also being developed at the NRC-HAA. The system will be integrated with the new readout under development in the coming months. The 850\,GHz module mechanical components are being machined and assembled at the NRC-HAA. Silicon AR lenses and metal mesh filters are being developed through contracts with U.\ Chicago and Cardiff respectively. The final 850\,GHz module will undergo various tests before being shipped to Cornell for integration in the Prime-Cam cryostat in $\approx$2024.


\acknowledgments 
 
SCC, AIH, AKS, JD, and JT, were  supported by the Canadian Foundation for Innovation and NSERC.
EV was supported by the NSF GRFP under Grant No. DGE-1650441. CCAT-prime funding has been provided by Cornell University, the Fred M. Young Jr. Charitable Fund, the German Research Foundation (DFG) through grant number INST 216/733-1 FUGG,  the Univ.~of Cologne, the Univ.~of Bonn, and the Canadian Atacama Telescope Consortium. MDN acknowledges support from NSF award AST-1454881.

\bibliographystyle{spiebib} 
\vskip1cm

\noindent {\bf REFERENCES}

[1] Niemack, M. D., “Designs for a large-aperture telescope to map the CMB 10x faster,” Applied Optics 55, 1686 (2016).

[2] Bustos, R., Rubio, M., Ot{\'a}rola, A., and Nagar, N., “Parque Astron{\'o}mico de Atacama: An ideal site for millimeter, submillimeter, and mid-infrared astronomy,” Publications of the Astronomical Society of the Pacific 946, 126 (2014).

[3] Stacey, G., et al., ``CCAT-Prime: science with an ultra-widefield submillimeter observatory on Cerro Chajnantor.'' SPIE 10700, 
    arXiv:1807.04354 (2018).

[4] Parshley, S. and et al., “CCAT-prime: The Fred Young Submillimeter Telescope Final Design and Fabrication,” Paper No. 12182-53 (2022).

[5] Vavagiakis, E. M., et al.\ 
“Prime-Cam: a first- light instrument for the CCAT-prime telescope,” in [Millimeter, Submillimeter, and Far-Infrared Detectors and Instrumentation for Astronomy IX], Zmuidzinas, J. and Gao, J.-R., eds., Society of Photo-Optical Instrumentation Engineers (SPIE) Conference Series 10708, 107081U (2018).

[6] Galitzki, N. et al., “The Simons Observatory cryogenic cameras,” Proc. SPIE (10708-3) (2018).

[7] Zhu, N. et al., “Simons Observatory large aperture telescope receiver design overview,” Proc. SPIE (10708-79) (2018).

[8] Vavagiakis, E. M., et al., ``CCAT-prime: Design of the Mod-Cam receiver and 280 GHz MKID instrument module'' SPIE,  arXiv~2208.05468, (2022).

[9] CCAT-Prime collaboration, et al., ``CCAT-prime Collaboration: Science Goals and Forecasts with Prime-Cam on the Fred Young Submillimeter Telescope'', ApJS submitted, arXiv 2107.10364 (2021).

[10] Radford, S. J. E. and Peterson, J. B., “Submillimeter Atmospheric Transparency at Mauna Kea, at the
South Pole, and at Chajnantor,” Pub. Astron. Soc. Pacific 128, 075001 (2016).

[11] Stacey, G., et al.\ ``SWCam: the short wavelength camera for the CCAT Observatory'' SPIE, Volume 9153, id. 91530L (2014).  

[12] Magnelli, B. et al.\ ``The deepest Herschel-PACS far-infrared survey: number counts and infrared luminosity functions from combined PEP/GOODS-H observations
'', Astronomy and Astrophysics, 553
132, (2013).

[13] Chapman, S., ``A Redshift Survey of the Submillimeter Galaxy Population'', ApJ 622 772 (2005).

[14] Pattle, K., \& Fissel, L., ``Magnetic fields in star formation: from clouds to cores'', Frontiers in Astronomy and
Space Sciences, 6, 15 (2019).

[15] Dober, B., et al.\ ``Optical Demonstration of THz, Dual- Polarization Sensitive Microwave Kinetic Inductance Detectors'', Journal of Low Temperature Physics,  https://arxiv.org/abs/1603.02963, (2016).

[16] Fischer, W.J., et al., ``Accretion Variability as a Guide to Stellar Mass Assembly'', Protostars and Planets VII, arXiv 2203.11257 (2022).

[17] Johnstone D., et al., ``Continuum Variability of Deeply Embedded Protostars as a Probe of Envelope Structure''
ApJ 765 133 (2013).

[18] Herczeg et al., ``How Do Stars Gain Their Mass? A JCMT/SCUBA-2 Transient Survey of Protostars in Nearby Star-forming Regions'',  ApJ 849 43 (2017).

[19] Lee, Y.-H., et al., ``The JCMT Transient Survey: Four-year Summary of Monitoring the Submillimeter Variability of Protostars'', ApJ 920 119 (2021).

[20] Geach, J., et al., ``The SCUBA-2 Cosmology Legacy Survey: 850um maps, catalogues and number counts'',
MNRAS 465 1789 (2017).

[21] Wang, W.H., et al.\ ``SCUBA-2 Ultra Deep Imaging EAO Survey (STUDIES): Faint-end Counts at 450um'',
ApJ, 850, 37 (2017).

[22] Geach, J., et al., ``The SCUBA-2 Cosmology Legacy Survey: blank-field number counts of 450-$\mu$m-selected galaxies and their contribution to the cosmic infrared background''
MNRAS, 432, 53 (2013).

[23] Parshley, S., et al., ``CCAT-prime: a novel telescope for sub-millimeter astronomy'',
SPIE 10700-5, (2018).

[24] Huber, A., et al., ``CCAT-prime: Optical and Mechanical design of the 850\,GHz camera module.'' Paper No. 12190-93 (2022).

[25] Griffin, J. Bock, and W. Gear, "Relative performance of filled and feedhorn-coupled focal-plane architectures," Appl. Opt.  vol. 41, no. 11, pp. 6543-6554, (2002).

[26] Sinclair, A. et al., “CCAT-prime: RFSoC based readout for frequency multiplexed kinetic inductance detectors,” Paper No. 12190-66,
arXiv~2208.07465, (2022).

[27] Austermann, J., Beall, J., Bryan, S. A., Dober, B., Gao, J., Hilton, G., Hubmayr, J., Mauskopf, P., McKenney, C., Simon, S. M., Ullom, J., Vissers, M., and Wilson, G. W., “Large format arrays of kinetic inductance detectors for the TolTEC millimeter-wave imaging polarimeter (Conference Presentation),” in [Millimeter, Submillimeter, and Far-Infrared Detectors and Instrumentation for Astronomy IX], 10708, 107080U, International Society for Optics and Photonics (2018).

[28] Choi, S.K., Duell, C.J., Austermann, J. et al. CCAT-Prime: Characterization of the First 280 GHz MKID Array for Prime-Cam. J Low Temp Phys (2022). https://doi.org/10.1007/s10909-022-02787-9

[29] X. Liu, W. Guo, Y. Wang, M. Dai, L. F. Wei, B. Dober, C. M. McKenney, G. C. Hilton, J. Hubmayr, J. E. Austermann, J. N. Ullom, J. Gao, and M. R. Vissers, "Superconducting micro-resonator arrays with ideal frequency spacing", Appl. Phys. Lett. 111, 252601, https://doi.org/10.1063/1.5016190,
(2017).

[30] Duell, C. et al., ``CCAT-prime: Designs and status of the first light 280 GHz MKID array and Mod-Cam receiver'', SPIE Paper Number: 11453-58  https://arxiv.org/abs/2012.10411 (2020).

[31] Britton, J.\ W.\, et al., ``Progress Toward Corrugated Feed Horn Arrays in Silicon'', 
Low Temperature Devices, 1185, (2009).

[32] Dobbs, M., et al., 
``Frequency multiplexed superconducting quantum interference device readout of large bolometer arrays for cosmic microwave background measurements'', RScI, 83g3113D, 
    arXiv:1112.4215  (2012).

[33] Rotermund, K., et al., ``Planar Lithographed Superconducting LC Resonators for Frequency-Domain Multiplexed Readout Systems'',  Journal of Low Temperature Physics, 184, page 486 (2016).

[34] Gordon, S., et al.\ 
“An open source, fpga-based lekid readout for blast-tng: Pre-flight results,” Journal of Astronomical Instrumentation 05(04), 1641003 (2016).

[35] Datta, R., Munson, C. D., Niemack, M. D., et al., “Large-aperture wide-bandwidth antireflection-coated
silicon lenses for millimeter wavelengths,” Applied Optics 52, 8747 (2013).

[36] Coppi, G. et al., “Cooldown strategies and transient thermal simulations for the Simons Observatory,” Proc. SPIE (10708-77) (2018).

[37] Orlowski-Scherer et al.\ “Simons Observatory large aperture receiver simulation overview ,” Proc. SPIE (10708-132) (2018).

\end{document}